\documentclass[twocolumn,pre,showpacs]{revtex4}

\usepackage{graphicx}
\usepackage{rotating}
\usepackage{amsmath}
\usepackage{amsfonts}
\usepackage{amssymb}
\usepackage{enumerate}
\usepackage{longtable}
\usepackage{dcolumn}
\usepackage{bm}
\usepackage{epsfig}
\usepackage{color}
\begin{document}

\title{Status maximization as a source of fairness in a networked dictator game}


\author{Jan E. Snellman$^{1}$}
\author{Gerardo I\~{n}iguez$^{4,2,1,3}$}
\author{J\'{a}nos Kert\'{e}sz$^{4,1}$}
\author{R. A. Barrio$^{5}$}
\author{Kimmo K. Kaski$^{1}$}
\affiliation{$^1$Department of Computer Science, Aalto University School of Science, FI-00076 AALTO, Finland}
\affiliation{$^2$Next Games, FI-00100 Helsinki, Finland}
\affiliation{$^3$Instituto de Investigaciones en Matem{\'a}ticas Aplicadas y en Sistemas, Universidad Nacional Aut{\'o}noma de  M{\'e}xico, 01000 M{\'e}xico D.F., Mexico}
\affiliation{$^4$Department of Network and Data Science, Central European University, H-1051 Budapest, Hungary}
\affiliation{$^5$Instituto de F{\'i}sica, Universidad Nacional Aut{\'o}noma de  M{\'e}xico, 01000 M{\'e}xico D.F., Mexico}

\date{\textrm{\today}}


\begin{abstract}
{Human behavioural patterns exhibit selfish or competitive, as well as selfless or altruistic tendencies, both of which have demonstrable effects on human social and economic activity. In behavioural economics, such effects have traditionally been illustrated experimentally via simple games like the dictator and ultimatum games. Experiments with these games suggest that, beyond rational economic thinking, human decision-making processes are influenced by social preferences, such as an inclination to fairness. In this study we suggest that the apparent gap between competitive and altruistic human tendencies can be bridged by assuming that people are primarily maximising their status, i.e., a utility function different from simple profit maximisation. To this end we analyse a simple agent-based model, where individuals play the repeated dictator game in a social network they can modify. As model parameters we consider the living costs and the rate at which agents forget infractions by others. We find that individual strategies used in the game vary greatly, from selfish to selfless, and that both of the above parameters determine when individuals form complex and cohesive social networks. }
{Dictator game, agent-based social simulation, superiority maximization}
\end{abstract}
\maketitle
%
%

\section{Introduction}
\label{intro}

People's behaviour towards others lies in a broad range from selfish to selfless: In the former, individuals are concerned with their own narrow minded benefit like profit, thus acting competitively against others, while in the latter they are concerned with the needs of the rest, that is, they behave in an altruistic and even self-sacrificing way. 

How people behave in real-life situations depends on the social or economic context and on individual characteristics. 
In order to get a deeper insight into human behavioural patterns in an economic context, Kahneman at al. designed the dictator game~\cite{KKT1986}, a two-player game where one of the players, the dictator, is tasked to divide a given sum of money between both players. The dictator can divide the money in any way, even keeping it all. The game was designed to test some of the assumptions of modern economic theories~\cite{HS}, mainly the assumption of rationality, under which the dictator will always keep all the money, not giving anything to the other player. In experiments, however, dictators tend to give at least a minor fraction of money to the other player, thus challenging the assumption of rationality of economic actors~\cite{FHSS1994,C2003,kysymys}. It should be emphasised at this point that the classical theory assumes that the utility function of the dictator is just personal profit.

The mismatch between theory and observation in the dictator game is largely caused by social factors, such as appreciation for fairness and equality, caring for others, reputation, etc. In other words, the empirical behaviour of the dictators may be understood as renouncing economic advantages to gain social status, such as esteem of the other player, or more generally a good reputation. While social status has been studied extensively in the social sciences, the analysis of its effects in economic theories (via, say, the concept of bounded rationality~\cite{HS} has been more limited). A way to bridge this gap between human social and economic behaviour is to consider how the wish to maximise social status may affect social network structure and thus the economic decisions of individuals. Social relations are a significant part of the overall social standing or status of people, with both negative and positive effects depending on the individual behaviour. Positive effects include increased  political and economic opportunities and social support in times of need, while negative effects can range from social exclusion to outright hostility. Hence humans usually have to take into account how their actions affect their relationships with others. In the context of the dictator game, we could see dictators weighing the worth of the monetary reward against the penalty incurred to their social relations, should they be seen to act too selfishly. 

The idea of status maximisation as a driving motivation of humans can be traced back to Adler's school of thought of individual psychology~\cite{adler}. Adler explored the scenario in which many psychological problems are the result of feelings of inferiority and, consequently, applied this idea to develop therapeutic techniques for what he termed inferiority complex. More recently, we used status maximisation (referred here as 'better than hypothesis' [BTH]) as a key mechanism to model the co-evolution of opinions and groups in social networks~\cite{SGGBK2017}. The aim of this study is to understand how a simulated society of individuals, driven by status maximisation, behaves when agents are allowed to freely form economic relationships but with a utility, where the profit aspect is only one component of status maximisation. To that end we use the BTH to model the social interactions of agents and their strategies in the dictator game. Previous agent-based models have considered the effect of social preferences such as inequality aversion and a tendency to fairness~\cite{X2010} but not status maximisation. Here we show how the drive of individuals to increase their status with respect to others determines both their economic strategies and the structural properties of the social network in which they reside.

Our approach to create a new utility function based on social status could be considered as an attempt to restore the idea of rational decision making in the economic interactions. However, we should emphasize that this interpretation has its limits. First, quantifying social status is highly non-trivial, but it could be circumvented by making simple assumptions. Second, while profit is simply measured in dollars (though the value of 1\$ is quite different for a beggar or for a millionaire), estimating social status has always a subjective component and it changes depending on the circumstances. Our purpose with the agent based approach is to test to what extent people are willing to invest in enhancing their social capital at the expense of their profits if their comparative position in the competition becomes better as a result.

The original dictator game is a two player game, where one of the players, the dictator, is tasked with dividing a fixed reward between the players  at will \cite{KKT1986,FHSS1994,C2003}. It is different from
the ultimatum game introduced in \cite{GSS1982}, where the second player gets to either accept or reject the offer, such that the rejection would result in neither player getting anything. Both games were constructed to demonstrate the limitations of rational economic behaviour, but in this study we focus
on the dictator game. 

Assuming economic rationality (or profit maximisation) on the part of the dictators one would expect them to keep the whole reward, and not give anything to the other player. However, in the hundreds of dictator game experiments in the past few decades it has been shown that many players, in the role of the dictator, actually do give out non-zero proportions of the reward to the other player (see e.g. the meta-analyses in \cite{C2003,CC2008,E2011}), which naturally challenges the notion of humans as rational economic actors. While there is a lot of variation in the results, the average offers to other player can reach as high proportions as almost $40\%$, as can be seen in Table \ref{tab1} for various studies, 
The effects of different social influences on the behaviour of the dictators have been studied extensively, (see for example, \cite{B2008} and references therein), and especially those of social norms (see e.g. \cite{RM2014,HEMBBBCGGHLMTZ2010,ZFS2015}). More recently, the effects of social networks, the players are embedded in, has been studied in \cite{BG2013} for the dictator game and in \cite{HCBWZGS2018} for the ultimatum game.

From the theoretical point of view, the experimental results obtained with the dictator game and similar games have been interpreted in terms of altruism \cite{L1998,CC2008}, fairness, in the sense of inequity aversion \cite{BO1998,BO2000,ES1999}, biological evolution \cite{A1981,SSM2015,AB2011}, or even experimental artefacts \cite{B2008,L2007}. Here we study the dictator game by means of agent based modelling in the context of coevolving social networks populated by agents, who are driven by maximising their social status. 

\begin{table*}[ht!]
\begin{tabular}{l c c c} 
Study & Type & Focus & Average offers \\
\hline
Engels 2011  & meta-analysis & general overview  & $28\%$ \\
Cardenas \& Carpenter 2008  & meta-analysis &  development & $34\%$ \\
Camerer 2003  & meta-analysis & game theory  & $\backsim 20\%$ \\
Zhao 2015  & experiment & politeness  & $39\%$ \\
Forsythe et al 1994  & experiment & real rewards  & $\backsim 20\%$ \\
\end{tabular}
\caption{Some of the more signicant studies of the dictator game summarized.} \label{tab1} 
\end{table*}

The paper is organized as follows: In the next section we introduce the networked dictator game model and explain the utility function we use for the evolution dynamics of the game, as well as the model parameters. Section \ref{resu} contains the results, including the analysis of the network geometries as a function of the parameters. In section \ref{con} we draw conclusions. 

\section{The network dictator game model}

Our dictator game model consists of a network of $N$ agents that form and break social connections to each other and play the dictator game repeatedly with the connected agents. The simulation proceeds in cycles, in which each connected agent plays the dictator game with all the agents connected to it. Note that for each pair of linked agents the game is played twice in each game cycle  with the agents exchanging the roles of the dictator and the supplicant between these instances of the game.

Each agent $i$ is characterised by the accumulated ``winnings'', or wealth, denoted by $v_i$. For every transaction of the game an amount $M$ of money is given to the agent acting as the dictator with 
its own dictatorial division strategy, denoted by $x_i$ ($0 \leq x_i \leq 1$) which 
is the proportion of $M$ that it gives to the other players. After each cycle is completed, an amount $cM$ is subtracted from every agent's wealth as living cost.  For the sake of simplicity we here 
assume that the proportion $c$ is the same for all the agents and for all times. If the wealth of an agent should go below zero, it is set to zero, meaning that the agent is still in the game. In addition to their social connections, the agents also adjust their division strategies during the simulation with 
a hill climbing algorithm,explained in more detail below.

When the agent $i$ plays with agent $j$, with $i$ being the dictator, the accumulated monetary reserves of the agents $i$ and $j$ change by 

\begin{eqnarray}
v_i(t_1) &=& v_i(t_0) + (1 - x_i) M, \\
v_j(t_1) &=& v_j(t_0) + x_i M,
\end{eqnarray}
where $t_0$ denotes the moment of time before the transaction and $t_1$ moment of time after the transaction. When one takes into account the reduction by $cM$, the full amount of accumulated wealth 
of the agent $i$ at cycle $T_1$ is as follows
\begin{equation}
v_i(T_1) = \\  \max\Big(v_i(T_0) 
+ M\big( k_i (1 - x_i) + \sum_{j \in m_i(1)} x_j - c \big) , 0 \Big),
\end{equation}
where $k_i$ is the number of neighbours of the agent $i$ and $T_0$ is the cycle preceding $T_1$.

The social network of agents was initially set to be randomly connected the average degree $\langle k \rangle$, but in the course of simulations there are no limitations on the degree of the agents. We deal with an adaptive network \cite{Kozma_Barrat2008,Iniguez_etal2009}: At the end of each cycle the network is let to reconfigure through rewiring the connections in such a way that the agents keep track of how the other agents affect their status. Here  we assume that the agents compare themselves against their neighbours, which means that every agent stores information not only what the other agents have given to it, but also what was given to its neighbours and how well they have accumulated wealth in comparison to them. For the sake of encouraging agents to renew their connections after negative experiences, we let their memories fade over time. 

The way the agent $i$ determines the influence of agent $j$ on its overall status is ultimately derived according to the BTH change of utility $\Delta_i$ of agent $i$. We assume that the agents only compare themselves to their neighbours (or, in other words, that the agents only know the accumulated wealth of their neighbours). 
Thus we can write the utility as follows 
\begin{eqnarray}
\Delta_i(T_1) &=&  v_i(T_1) - v_i(T_0) + \sum_{l \in m_1(i)} (v_i(T_1) - v_l(T_1)) \nonumber \\
 && - \sum_{l \in m_1(i)} (v_i(T_0) - v_l(T_0)).
\label{deq}
\end{eqnarray}
To see how the actions of agent $j$ affect $\Delta_i$ it is instructive to rearrange the terms of $\Delta_i$ in the following way:
\begin{eqnarray}
\Delta_i(T_1) 
&=& (k_i(T_0) + 1) (v_i(T_1) - v_i(T_0)) \nonumber \\
&& - \sum_{l \in m_1(i)} (v_l(T_1) - v_l(T_0)) \nonumber \\
&=& (k_i(T_0) + 1) (v_i(T_1) - v_i(T_0)) \nonumber \\
&& - \sum_{l \in m_1(i)/\{j\}} (v_l(T_1) - v_l(T_0)) \nonumber \\
&& - (v_j(T_1) - v_j(T_0)), 
\label{uijhelpeq}
\end{eqnarray}
where $k_i$ is the degree of agent $i$. From Eq. \ref{uijhelpeq} one can see that agent $j$ can influence $\Delta_i$ in three ways: first by giving money to $i$, second by giving money to the other neighbours of $i$ and third by accumulating money itself. The amount of money given by agent $j$ to all connected agents is $x_j M$ per dictator game cycle. Thus, it is possible to define a cumulative utility matrix $U_{ij}$ to describe how the agent $i$ has benefited from the actions of agent $j$ 
at cycle $T_1$ as 

\begin{equation}
U_{ij}(T_1) = U_{ij}(T_0) + a_{ij} U'_{ij}(T_1) + \gamma_{ij}(T_0),
\label{uijeq}
\end{equation}
where $a_{ij}$ is an element of the adjacency matrix:
\begin{equation}
a_{ij} = 
\left\{
\begin{aligned}
&1,\;\; \mathrm{if}\;\mathrm{agents}\;\mathrm{i}\;\mathrm{and}\; \mathrm{j}\; \mathrm{are}\;\mathrm{linked}\;\;\\
&0,\;\; \mathrm{otherwise},
\end{aligned}
\right.
\end{equation}
$\gamma_{ij}$ is the matrix of memory parameters, $n_I$ is the amount of agents in $I = m_1(i) \cap m_1(j)$ and 
\begin{eqnarray}
U'_{ij}(T_1) 
&=& (k_i(T_0) -n_I + 1)x_j(T_0) M  \nonumber \\
&& -  (v_j(T_1) - v_j(T_0))
\label{uprime}
\end{eqnarray}
The memory parameter matrix $\gamma_{ij}$ measures the speed at which the agents forget how they were treated, and it is designed so as to reduce $\lvert U_{ij} \lvert$ to zero in time. Thus, it can be written in the form
\begin{equation}
\label{gamma0}
\gamma_{ij}(T_0) = 
\left\{
\begin{aligned}
&\gamma_{0},\;\; \mathrm{if}\;\; U^T_{ij}(T_0) \ge \gamma_{0},\\
&-\gamma_{0},\;\; \mathrm{if}\;\; U^T_{ij}(T_0) \le -\gamma_{0}\\
&U^T_{ij}(T_0),\;\; \mathrm{if}\;\;  -\gamma_{0} \le U^T_{ij}(T_0) \le 0\\
&-U^T_{ij}(T_0),\;\; \mathrm{if}\;\;  0 \le U^T_{ij}(T_0) \le \gamma_{0}
\end{aligned}
\right.
\end{equation}
where the memory parameter $\gamma_{0}$ (assumed to be constant for the sake of simplicity) is the maximum pace of forgetting and $U^T_{ij}(T_0) = U_{ij}(T_0) + a_{ij} U'_{ij}(T_1)$. For the sake of simplicity we also assume that the agents have full knowledge where the accumulated wealth of the other agents is coming from, so, for instance, the agent $i$ can adjust $U_{ij}$ even if the agent $j$ is not connected to it.

After each cycle of the game, the agents form connections with the agents that have benefited them and cut connections with agents that have not, in other words the agent $i$ will form a connection with the agent $j$ if such a link is not already present and $U_{ij} \ge 0$, and cut an existing link with the agent $j$ if $U_{ij} < 0$.

The agents adjust their division strategies $x_i$ using a simple hill climbing algorithm. At first, the $x_i$'s are randomized, and changed by $dx$ at every step. In the second step, $dx$ is randomly chosen to be either $-0.1$ or $0.1$. According to BTH, the change of status of the agent $i$ in between game cycles is given by Eq. (\ref{deq}), which determines the further evolution of $x_i$: if $\Delta_i \ge 0$, the direction of $dx$ is the same as before, but  if $\Delta_i < 0$, the direction of $dx$ is changed. Inspired by the simulated annealing techniques, we reduce $\lvert dx \lvert$ linearly during the first $1000$ time steps to a minimum of $0.01$. 

\subsection{Motivation of the model parameters $c$ and $\gamma_{0}$}

The main motivation for including the cost parameter $c$ in the dictator game model is to test the effect of mutual dependence on the social systems of the agents. The cost parameter is important from the BTH (Better Than Hypothesis) perspective, because it can be interpreted to represent outside pressure to the agents. Under the BTH assumption it is not immediately clear whether the status maximizing agents would form social bonds of any type, let alone for the purpose of playing the dictator game. However, it is conceivable that some common needs might force the agents to interact socially. Thus, reducing the wealths of the agents by an amount controlled by the parameter $c$, introduces into the model an effect that requires agents to cooperate in order to gain anything in the long run. This then allows us testing whether mutual need enhances social interactions between the agents. 

The potential range of the cost parameter $c$ can in principle extend to be positive or negative without limits, but will be limited for the purposes of this study by considering the effects of the parameter to the wealths of the agents. In order for an agent to make profit in the model it needs to have a sufficient number of neighbours that are willing to play the dictator game with them on good enough terms. When $c = 0$, the agents retain all the wealth they manage to acquire from their dictator game interactions with others for all time, while if $c > 0$, their wealths slowly decline if not replenished through the dictator game. As a direct consequence of these facts the agents need more and more neighbours to be able to cover their expenses as the parameter $c$ is increased. If the agent $i$ has only one neighbour, $j$, it can generate profit from their relationship as long as $c < 2$, if it uses totally selfish strategy ($x_i = 0$) and its partner is totally generous ($x_j = 1$). Of course, this arrangement is disadvantageous to the agent $j$, and therefore not likely to happen, unless agent $j$ happens to have a multitude of other more generous neighbours. A pair of agents using similar division strategies can only make profit if $c < 1$, but when $c$ is increased, an agent needs at least $\lceil c \rceil$ neighbours with similar division strategy to cover the costs of its living standard. 

As can be seen, the theoretical maximum profit an agent can make from one relationship per simulation cycle is $2M$, while more realistically it can be expected to amount to about $M$. In any case, when $c$ is high enough, the agents cannot cover their costs anymore even if they form social links with every other agent in the simulation. In a simulation with $N$ agents this point can be expected to be reached at the latest somewhere between $c = N - 1$ and $c = 2(N - 1)$, depending on the configuration of the social network of the agents and their division strategies.   In this study we do not look into the effects of ``universal basic income'', i.e. the $c < 0$ case, and we limit our scrutiny of the cost parameter well below the upper limit of $2(N - 1)$.  

The function of the memory parameter $\gamma_{0}$ in the dictator game model is to allow the agents to reform links that have once been broken, ensuring the continuation of the social dynamics.  
Without the moderating influence of the memory mechanics in the model, the space of possible social connections would steadily diminish during the model simulations, resulting in a very limited social network. 

The interesting range of
the memory parameter $\gamma_{0}$ can be estimated using the same procedure as the one used for the cost parameter $c$, i.e. by calculating the point at which the parameter's influence overwhelms everything else. Negative values for $\gamma_{0}$ would make no sense in our context, so the lower limit can be set to $0$. The maximum limit can be estimated using Eq. (\ref{uijeq}), from which it can be seen that in the case of the memory parameter finding this limit amounts to finding the maximum value of $\vert U'_{ij}(T_1) \lvert$, which $\gamma_{0}$ would need to exceed. From the definition given in Eq. (\ref{uprime}), we can see that $U'_{ij}(T_1)$ depends in a rather complicated way on both the structure of the social network of the agents and their division strategies, but thankfully there are only two terms to consider. Let us denote these terms as 
\begin{eqnarray}
a &=& (k_i(T_0) -n_I + 1)x_j(T_0) M \\
b &=& v_j(T_1) - v_j(T_0),
\end{eqnarray}
so that $U'_{ij}(T_1) = a - b$. Since necessarily $k_i(T_0) \geq n_I$, it follows that $a \geq 0$ always. 

The term $a$ attains its minimum value  of $0$ when $x_j(T_0) = 0$, and its maximum value of $NM$ when $k_i(T_0) = N - 1$, $n_I = 0$ and $x_j(T_0) = 1$, i.e. when the agent $i$ is connected to all other agents and the agent $j$ has no other connections and uses the most generous strategy possible in the dictator game. Similarly, the term $b$ has a minimum value of $-cM$ when the agent $j$ receives nothing from the other agents, and a maximum value of $(2(N-1) -c)M$ when $x_j(T_0) = 0$ and $x_k(T_0) = 1$ for all $k \neq j$. As can be seen, the maximum value for $a$ can occur simultaneously with the minimum value of $b$ and vice versa, which means that the maximum value of $\vert U'_{ij}(T_1) \lvert$ can be found  
either in the case where the term $a$ is at maximum and $b$ at minimum or in the case where $a$ is at minimum and $b$ at maximum. The latter of these cases yields the greater value for $\vert U'_{ij}(T_1) \lvert$, amounting to a total of $(2(N-1) -c)M$. This is therefore a sensible upper limit to $\gamma_{0}$, since beyond that one would expect the social dynamics to settle. As in the case of the cost parameter, we limit our study to relatively small values of $\gamma_0$, so we do not 
approach the upper limit $(2(N-1) -c)M$.

\section{Results}
\label{resu}

We initialise the dictator game model with $N$ agents each having a random dictatorial strategy or proportion of the total amount, $x_i$, the agent gives to the other player. In the simulation run at each time step each one of the $N$ agents in turn acts as a dictator and we let the system to run for a fixed period of $10000$ time steps. For $M$ we use the value of $1$. In this work we focused on investigating the following characteristic network quantities, i.e. the average degree $\langle k \rangle$, the average shortest path $\langle L \rangle$, the  local and average clustering coefficients $C_i$ and $\langle C \rangle$ , the mean number of second neighbours $\langle n^{(2)} \rangle$, the average assortativity coefficient $\langle r_a \rangle$, and the average homophily coefficient $\langle r_h \rangle$. 
In addition, we measure the susceptibility $\langle s \rangle$, which is the second moment of the number of $s$ sized clusters, $n_s$:
\begin{equation}
\langle s \rangle = \frac{\sum_s n_s s^2 }{\sum_s n_s s}.
\label{suscep}
\end{equation}
As in percolation theory, the contribution of the largest connected component 
of the network is neglected when calculating the susceptibility (\ref{suscep}). Furthermore, we investigated the assortativity and homophily coefficients defined as the Pearson correlation coefficients of the degrees $k$ and accumulated wealth $v$ of linked agents, respectively, as discussed in \cite{PPZ} and \cite{SGGBK2017}. It should be noted that the assortativity and homophily coefficients are ill defined if agents all have exactly the same amount of neighbours and if they are all connected to agents with exactly the same amount of wealth, respectively. These situations do rise in our simulations occasionally, and when they do, the results for the assortativity and homophily coefficients are excluded from averages calculated. We also calculated  the Gini coefficient as a well known measure of inequality, first proposed by Gini in 1912 \cite{gini1912}, using the following definition 

\begin{equation}
G = \frac{\sum_i^N \sum_j^N \lvert v_i - v_j \lvert }{ 2 N \sum_i v_i},
\label{ginicoef}
\end{equation}
which basically measures the total difference between the accumulated wealths $v$ of agents. 

\subsection{Time-evolution of the model}

In order to obtain sufficiently good statistics for determining the averages of the quantities listed above the simulations of the model were  run for $100$ realisations, and time averages over the latter half of the time series were also taken. The reason for taking the time averages from the realisations was the very fluctuating nature of the time-evolution of the model. At times, the entire social network may cease to exist temporarily, although these moments only occur within certain ranges of the model parameters, especially when large values of $\gamma_0$ are involved.

In order to study the time-evolution of the properties of the agents and their social networks in our model we performed two singular simulations with $N = 100$ and two different sets of parameter values, the first set being $\gamma_0 = 0$ and $c = 5$ (case A), and the second $\gamma_0 = 5$ and $c = 0$ (case B). In addition to determining the proper measure for the averages of simulation results,  the main motivations for these experiments were to test converge on the other hand, and to see how the model parameters influence the temporal behaviour of the model on the other. For example, one could surmise from the very definition that $\gamma_0$ could potentially have major effects on the time-evolution of the social networks of the modeled agents. 

\begin{figure}[h]
\begin{center}
\epsfxsize = 0.73\columnwidth \epsffile{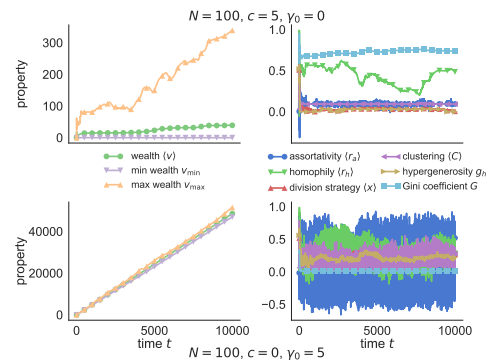}
\caption{The time-evolution of the agent and network properties when $N = 100$, with $\gamma_0 = 0$ and $c = 5$ (above), or $\gamma_0 = 5$ and $c = 0$ (below). The minimum, average and maximum wealths of agents are displayed in the panels on the left, while the right panels show the hypergenerosity prevalence along with homophily, assortativity, clustering and gini coefficients.}
\label{timec50m50}
\end{center}
\end{figure}

The results are illustrated in Fig. \ref{timec50m50}. The network properties seem to generally converge to some constant values around which they fluctuate, but in case B these fluctuations are very strong. The assortativity coefficient especially becomes almost meaningless as it can have both negative and positive values in a very short period of time due to the fluctuations, implying that the agents have no clear preference on whether to seek connections with those of same or different degree. In case A, in contrast, while there are still relatively large fluctuations in the value of the assortativity coefficient, the overall value of the coefficient is clearly positive. While the fluctuations of most network properties are very rapid, the homophily coefficient in case A exhibits slowly varying behaviour. 

In contrast to the network properties, the time evolution of the minimum, maximum and average monetary reserves of agents in the simulations show no signs of fast fluctuations. Also, it turns out that in this case it is the parameter $c$ that has greater impact: While in case B the growth of all of the reserves is almost linear, in case A the growth of the maximum and average reserves stall eventually and finally start fluctuating slowly around constant values, while the minimum reserve stays stubbornly at $0$. There is thus a substantial difference in behaviour between cases A and B when it comes to relative differences between the minimum, maximum and average monetary reserves of the agents. All the reserves reach tens of thousands in value in case B, while in case A they do not rise above 300. Also, in case A the maximum reserves are in the final stages of the simulation about four times the size of average reserves, while in case B all the reserves are on relative terms very close to each other, only diverging very slowly. These behavioural differences are reflected in the gini coefficient, as it tends to $0$ in case A, and to a value of little over $0.6$ in case B, owing to the more unequal wealth distribution in the latter case.

\subsection{General characteristics of the social networks produced by the model}

\begin{figure}[h]
\begin{center}
\epsfxsize = 0.73\columnwidth \epsffile{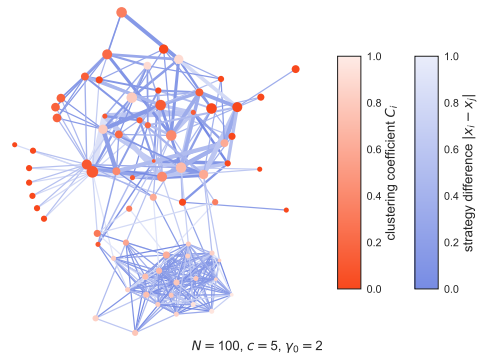}
\caption{An example of the final state of the social network produced by our model, when $N=100$, $c = 5$ and $\gamma_0 = 2$. The colours of the nodes show the local clustering coefficients, while the colour of the edges show the proximity of the division strategies employed by the agents linked by the edge, as indicated. The sizes of the nodes correspond to the accumulated wealths $v_i$ of the agents, while the widths of the edges correspond to the strength of the connection between linked agents, i.e $min( \{ U_{ij},U_{ji} \} )$. }
\label{netfig}
\end{center}
\end{figure}

The model parameters $c$ and $\gamma_0$ have a strong influence on the structure of the social network produced by the model and the division strategies of the agents. The social networks produced by the agents in the model can vary from very simple to very complex depending on the values chosen for these parameters. For a reference, a final state of the social network with $N = 100$ agents, $c = 5$ and $\gamma_0 = 2$ is shown in Fig. \ref{netfig}. With these parameter values the social networks of the agents produced by the model acquire their most complex form and exhibit clearly their most interesting features. Next we explain what kinds of simpler forms the network may take and with what parameter values, and how the complex network shown in Fig. \ref{netfig} emerges from these simpler forms. 

When the model parameters are $c = \gamma_0 = 0$, the networks formed consist only of collections of pairs or short chains of linked agents. The more agents a chain has, the rarer that chain is in the network. The total amount of agents in a simulation also determines how long the chains can get: chains longer than four agents seem to never occur in simulations of $100$ agents, but the chains of even nine agents can manifest when the total population in the simulation is increased to $300$ agents. The strategies $x_i$ employed by the linked agents are invariably most generous possible, that is $x_i = 1$. This is most likely due to lack of a reason for the agents to tolerate unfairness when $c = 0$, while the disconnected and linear nature of the social networks formed by the agents is probably due to the unforgiving nature of relation formation when  $\gamma_0 = 0$. 

Increasing the parameter $\gamma_0$ from the value of $0$ causes the network structure to become gradually more complicated.While the social networks of agents are generally disjointed at relatively small values of $\gamma_0$, the relatively short chains of agents lengthen as $\gamma_0$ increases, and at large values of $\gamma_0$ these chains become tangled. As $\gamma_0$ is increased further, this tangledness only increases, until the networks closely resemble that shown in Fig. \ref{netfig}. At some point the agents start forming densely connected hubs within the larger networks, which we name as ``trade associations''. The agents in these associations seem to have relatively coordinated strategies in comparison to the agents outside these associations, which may be a result of different ``grand strategies'' utilized by these different types of agents. The agents within the trade associations seek fair exchange with other agents in the same association, which in the context of our model means having the same division strategies while the agents outside these associations generally fall into two categories: those using relatively generous offerings to attract many less generous partners, and those agents that in turn take advantage of the more generous agents, but have few other social connections themselves. In a way, this arrangement is reminiscent of the patron-client relationships,  and as such we call the more loosely connected part of the main component "patron-client network". As for the offering proportions $x_i$, the agents generally adopt less generous strategies as the memory parameter $\gamma_0$ is increased, except for the trade associations, whose strategies may be more flexible. In Fig. \ref{netfig} the network has a trade association in its lower part, while the rest is composed of a patron client network.

Features that emerge only occasionally in our simulations, but often enough to be noticeable, are totally connected components that are completely disconnected from the main network, and whose agents have totally convergent division strategies. Obviously, these formations are extreme cases of trade associations, and as such we call them ``cartels''. These cartels can be unstable in the sense that they may periodically disband and reform, but they may also be very robust at times. 

When the parameter $c$ is increased when $\gamma_0=0$, the strategies $x_i$ gradually become less generous, while the general structure of the social network remains initially the same as in the $c = \gamma_0 = 0$ case, i.e. disjointed collections of small chains of agents. However, when $c$ is increased sufficiently, the networks finally become more complex. The network structure in the case of large $c$ and no $\gamma_0$ shows some similarities to the one shown in Fig. \ref{netfig}, such as clusters of densely connected agents reminiscent of trade associations, but the distinction between these and the patron-client network is weaker. Furthermore, in the case of large $c$ the division strategies most often decline to zero for all linked agents. Thus, one cannot say for sure if any strategy coordination is taking place. One needs to remember that, while $\gamma_0$ has a direct effect on the relation formation behaviour of the agents, $c$ only has an indirect effect through the term $f=-(v_j(T_1) - v_j(T_0))$ in Eq. \ref{uprime}, and that while the former plays a  role both in the forming and breaking of relations, the effect of the latter turns out to be to purely discourage the breaking of the relations. This is simply because $c$ always contributes positively to the $U_{ij}$, since
\begin{equation}
\label{ceffect}
f = 
\left\{
\begin{aligned}
&(c - k_j(x_j - 1) - \sum_{a \in m_1(j)} x_a )M,\;\; \mathrm{if}\;\; c \ge c_{t},\\
&v_j(T_0),\;\; \mathrm{if}\;\; c \le c_{t},
\end{aligned}
\right.
\end{equation}
where 
\begin{equation}
c_{t} = v_j(T_0) + (k_j(x_j - 1) + \sum_{a \in m_1(j)} x_a )M.
\end{equation}
Thus, with sufficiently large $c$ and small average degrees $U_{ij}$ will always remain positive, and no existing relations are ever broken. Most likely the typical course of a simulation in $\gamma_0=0$ case is that first the all the agents in the simulation form a fully connected community at the first time step, and subsequently most of the links between the agents will be cut until the degrees of all agents with any connections left are below what are allowed by $c$, after which the network remains unchanged for the remainder of the simulation. 

The likely reason for the formation of chains of linked agents in the low $c$ and $\gamma_0$ cases is the fact that any offerings by an agent to the neighbours of its its neighbours weaken the standing of the said neighbours. The formation of the trade associations is probably connected to these associations becoming socially acceptable when forgiveness ($\gamma_0$) of agents allows, or when there is enough outside pressure ($c$) to the agents, or a suitable combination of these effects. Generally it seems as though $\gamma_0$ on its own has a stronger effect on the network structure than $c$, which on its own seems to have a stronger effect on the division strategies of agents than the network structure. 

The most relevant parameter values are those that result in the agents having similar division strategies to those found in the real world experiments on the dictator game. Since our model is too simple to reproduce the results of the experiments one-to-one, we focus only on the most features of these results that are most relevant in the context of our model. These features are the fact that, on one hand, surprisingly large proportion of dictators give something to the other player (according to \cite{E2011}, only about $36\%$ of dictators choose to give nothing, while the average giving rate is about $28\%$), and on the other hand the fact that the distribution of giving rates is strongly skewed in favour of the dictator, with only about $12\%$ of dictators giving more than $50\%$ of the reward to the other player according to \cite{E2011}. In this study we use the term hypergenerosity for the tendency of the agents to give more than $50\%$.

\begin{figure}[ht]
\begin{center}
\epsfxsize = 0.73\columnwidth \epsfysize = 0.33\textheight \epsffile{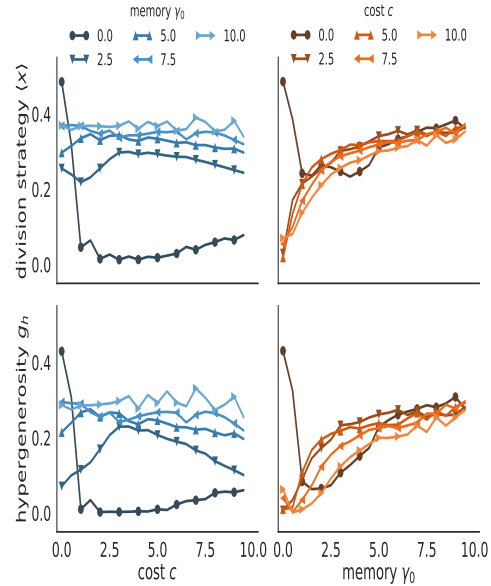}
\caption{The average division strategies (upper panels) and the prevalence of hypergenerosity (lower panels) as functions of $c$ with five constant values of $\gamma_0$ (panels on the left) and $\gamma_0$ with five constant values of $c$ (panels on the right).}
\label{hyperfigure}
\end{center}
\end{figure}

In order to find some values for the parameters with which our model is able to at least some extent match the average division strategies and the prevalence of hypergenerosity, latter of which is denoted here by $g_h$, we performed simulations in which either $c$ or $\gamma_0$ was kept constant, and the other was varied. The values tested were $0$, $2.5$, $5$, $7.5$ and $10$ for both parameters. As for a definition for hypergenerosity prevalence we simply adopt the proportion of agents with $x_i > 0.5$ of all agents.

\begin{figure}[h]
\begin{center}
\epsfxsize = 0.73\columnwidth \epsffile{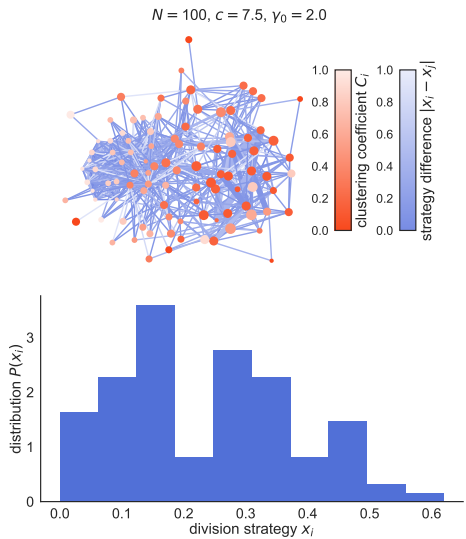}
\caption{An example of the social network, when $N=100$ and $c = 7.5$ and $\gamma_0 = 2.0$, shown in the upper panel. Colours, line widths, etc as indicated in Fig. \ref{netfig}. A histogram of the resulting division strategies is shown in the lower panel.}
\label{Cgfig}
\end{center}
\end{figure}

The results for average division strategies and hypergenerosity are shown in Fig. \ref{hyperfigure}. Considering the fact that our agents are driven by the motive of superiority maximization, hypergenerosity is surprisingly common in our simulations. Most notably, when one looks the lower left panel of Fig. \ref{hyperfigure}, one sees that for fixed values of $5.0$, $7.5$ and $10.0$ for $\gamma_0$, $g_h$ stays between values of approximately $0.2$ and $0.35$ for all values of $c$, which is considerably above the $12\%$ proportion reported in \cite{E2011}. Also, when $\gamma_0$ is given value $2.5$, $g_h$ only drops below $0.12$ when either $c \lesssim 1.5$ or $c \gtrsim 8.0$. The highest proportion of overgenerous agents, or about $42\%$, occurs when $\gamma_0 = c = 0$, but if $c$ is increased while $\gamma_0$ is kept constant, this proportion declines fast to a value only little over zero, as could be expected from the social network behaviours discussed above. While there is some rise with larger values of $c$, $g_h$ only rises to about $5\%$ at most in the very largest values of $c$ tested. In general, from the lower left panel of Fig. \ref{hyperfigure} one could draw the conclusion that the parameter $\gamma_0$ for the most part enhances $g_h$, an observation that is broadly confirmed by the lower right panel of Fig. \ref{hyperfigure}: For all the constant $c$ values shown the trends are growing, at least when $\gamma_0 \lesssim 1.5$ 

As for the division strategies, it is interesting to note that when $\gamma_0 > 0$ the parameter $c$ has relatively little influence to the strategies, which generally hover around or above values of $0.25$, $0.30$ or $0.35$, depending on the value of $\gamma_0$ chosen. The value $0.30$ happens to be relatively close to the value found in \cite{E2011}, while $0.35$ is closer to findings of \cite{CC2008}. Only in the $\gamma_0 = 0$ case we can see the strong drop in the generosity of the agents, a phenomenon already identified above in the context of the social networks. As a function of $\gamma_0$ the average division strategy tends to be generally increasing, except in the $c = 0$ case, in which it first declines from a high of almost $0.5$ to about $0.25$, stagnates there and then starts increasing. In all other tested cases the increasing trends are rapid at first, but gradually slow down as $\gamma_0$ increases. The values of the average division strategies approximately stay between the $28\%$ of \cite{E2011} and $34\%$ of \cite{CC2008} when $\gamma_0 \gtrsim 2$ for the $c = 5$ case. The same holds for the $c = 10$ case when $\gamma_0 \gtrsim 5$, and for the $c = 2.5$ and $c = 7.5$ cases when $\gamma_0 \gtrsim 4$.

\begin{figure}[h]
\begin{center}
\epsfxsize = 0.73\columnwidth \epsffile{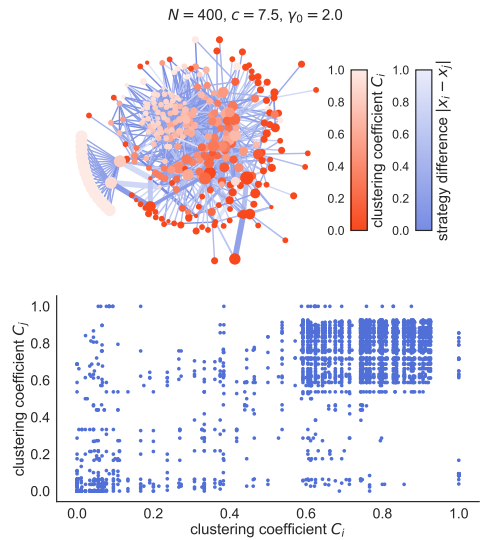}
\caption{A social network resulting from a run of $400$ agents with $c = 7.5$ and $\gamma_0 = 2.0$. A scatterplot of local clustering coefficients of linked agents are shown in the lower panel.}
\label{CCfigure}
\end{center}
\end{figure}

Choosing the most realistic values for the model parameters, i.e. those that yield the average hypergenosity prevalence and division strategies closest to those observed, involves a careful consideration of the effects of the parameter $\gamma_0$. On the one hand, this parameter increases hypergenerosity prevalence, so it should not be too high, but on the other hand it also increases average division strategies, so it should not be too low either. While this dilemma places tight restraints on realistic parameter values, Fig. \ref{hyperfigure} shows that with $\gamma_0 = 2$ and $c = 7.5$ $g_h$ is close to $12\%$ and an average division strategy is about $0.25$, which are very close to the ones observed in \cite{E2011}, although the division strategy is a little bit lower. Better match might be found with a more through sweep of the parameter space, but for our purposes this result is close enough. Besides, as shown in Table \ref{tab1} there is a great deal of variance between results of different studies, with some of the older results being as low as $20\%$. So, it could be argued that the $25\%$ is clearly within the spectrum of acceptable results. Fig. \ref{Cgfig} shows an example of the social network generated with $\gamma_0 = 2$ and $c = 7.5$, along with a histogram showing the frequency distribution of the division strategies. The network is somewhat to similar to the one shown in Fig. \ref{netfig}, with a trade association and a patron-client network, and relatively dense interconnections between these two components. 

The histogram in Fig \ref{Cgfig} shows a distribution of division strategies of a peculiar shape with at least three local peaks. The peak at about $x_i = 0.5$ corresponds to fair division, and such a peak has also been observed in experiments \cite{E2011}. The other two peaks at about $x_i = 0.15$ and $0.3$, however, have no counterparts in the experiments, and there is no peak at $x_i = 0$ (i.e. the point the agent playing as dictator keeps everything to itself) as one would expect from the experimental studies. That our simulated experiments do not correspond one to one to experimental observations is on one hand not surprising, since our setup differs greatly from typical dictator game experiments, but on the other hand the distribution of division strategies indicates that the special cases of totally fair ($x_i = 0.5$) and unfair ($x_i = 0$) divisions have no intrinsic meaning to our agents, a situation which might change if the agents were made to compete for social goodwill as well as accumulated wealths.  

While the social networks of the agents belonging to trade associations tend to be fully connected, those of patron-client networks are characterised by aversion to forming triangles. This can be seen by comparing the local clustering coefficients of linked agents, as in Fig. \ref{CCfigure}, in which we show an example of a larger run with $400$ agents, and in which $c = 7.5$ and $\gamma_0 = 2.0$. The links shown in the figure tend to concentrate in the upper right and lower left portions of the plot, which provides a visual presentation of the different parts of the full social network: In the trade association the agents tend to be all connected to each other, thus they all have high local clustering coefficients, and so the connections between them show up in the upper right portion of Fig. \ref{CCfigure}. The agents of the patron-client network have sparser connections and, therefore, lower local clustering coefficients, and since they also mostly connect with each other, their connections populate the lower left proportion of the plot. The social relations formed by the agents that connect these two communities are rarer, and show up outside these areas. Especially strong are the sets of points running vertically and horizontally in straight lines in the middle of the plot, which correspond to the links of the peripheral members of the trade association seen in the network. Since this particular network contains many agents with only a single neighbour, for whom local clustering coefficient is zero, and these particular agents tend to connect the more sparsely connected part of the network, there are especially prominent concentrations of points on the lower parts of the axes.

With the parameter values $c = 7.5$ and $\gamma_0 = 2.0$, our model yields results for average hypergenerosity prevalence and division strategies that very roughly correspond to those found experimentally. While parameters producing a better match might be found by combing the parameter space more carefully, it is also interesting to ask, what do these (or any other) values tell about the simulated society of  the agents, and what are their relevance to the real human societies? 

In the case of the parameter $c$ the answers to these questions are relatively straightforward: As noted above, $\lceil c \rceil$ is the minimum number of social connections one needs in order to make profit in the game, when similar division strategies are used. Thus $c = 7.5$ implies that maintenance of the life-style of a single person requires co-operation with at least eight other people, which suggests a significant degree of interdependence in the social system. The implications of the parameter $\gamma_0$ are harder to quantify, but in the simple system of only two agents $i$ and $j$ linked to each other $\gamma_0 = 2.0$ would allow the agents to continue interacting even in an extremely unfair setting, that is, for example, when $x_i = 1$ and $x_j = 0$. This may seem to indicate rather radical levels of tolerance of unfairness on part of the agents, but in the context of our simulations we had $100$ agents instead of only two, and the average degree of the agents was well in excess of eight. Since the components of the utility matrix $U_{ij}$ of the agents change much more rapidly in this situation, $\gamma_0 = 2.0$ is really not that high. Take, for example, a fully connected subgroup of nine members, where agent $1$ has a selfish division strategy $x_1 = 0$, while all the others have hypergenerous strategy $x_i = 1$. In this case, $U'_{1j} = -16$ after one round, which cannot be canceled by such a small $\gamma_0$. Thus we can say that the effect of $\gamma_0$ parameter is rather subtle in the context of our illustration case $c = 7.5$ and $\gamma_0 = 2.0$. 

\subsection{The properties of the network and wealth distribution}

\begin{figure}[h]
\begin{center}
\epsfxsize = 0.73\columnwidth \epsffile{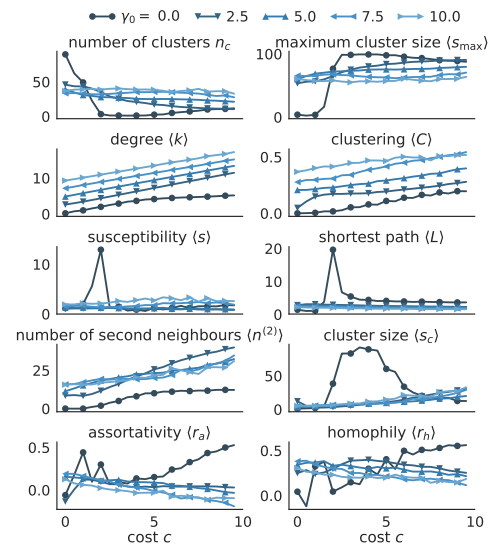}
\caption{The network properties as functions of $c$. The linestyles indicate, which of the five constant values of $\gamma_0$ is being shown.}
\label{ultcfig}
\end{center}
\end{figure}

In order to get more detailed picture on how our model behaves as functions of its parameters, we performed simulations to determine the averaged network properties mentioned above. Fig. \ref{ultcfig} shows these averaged network properties as functions of the parameter $c$, for $\gamma_0 = 0, 2.5, 5, 7.5, 10$. One of the most noticeable features seen in the Figure is that in the $\gamma_0 = 0$ case the behaviour of the network properties is in most cases very different to those of other $\gamma_0$ values, which tend to behave similarly to each other. Only the average numbers of first and second neighbours and clustering coefficient exhibit somewhat similar behaviours for all $\gamma_0$ values.

The behaviours discussed in the context of Fig. \ref{netfig} are readily apparent in the Fig. \ref{ultcfig}, especially in the $\gamma_0 = 0$ case. The steep decline in the number of clusters and the increases in the average and maximum cluster sizes in this case are the most obvious indication of the transition from the collections of short chains of agents to the more connected networks as $c$ increases, which is also evident in the steadily increasing average clustering coefficient. Average susceptibility and path length give interesting insight into the intermediate stages of this transition, as they both spike at $c = 2$, which is due to the main component of the social network becoming a one long chain or a loop. It would seem that at this point the amount of agents belonging to clusters outside the main component reaches a maximum, as does the length of the chains of agents in the main component.  

As stated above, the results for the network properties generally follow the same trends for all the other values of $\gamma_0$ tested. However, for all $\gamma_0$ the average numbers of first and second neighbours and the average clustering coefficient rise as $c$ increases, implying that the parameter $c$ has a universally enhancing effect on the connectedness of the agents. Especially the rising average clustering coefficient might be an indication of growing trade associations. As for the number of clusters and maximum cluster size, the results obtained using $\gamma_0$ values other than zero follow the trends expected from earlier analysis, as the former declines and latter increases, although these trends are not nearly as clear as in the $\gamma_0 = 0$ case, and for $\gamma_0 = 7.5$ and $\gamma_0 = 10$ the maximum cluster size is almost constant. Also, when $\gamma_0 \neq 0$ the average cluster size increases monotonically as a function of $c$, while for $c \lesssim 3.5$ and $\gamma_0 = 0$ it actually starts decreasing from its maximum value after a sharp increase, suggesting re-emergence of clusters outside the main component of the network. Unlike the $\gamma_0 = 0$ case, the susceptibility and the average path length do not exhibit drastic changes as functions of $c$ for other values of $\gamma_0$, although there may be slight increasing trend in the case of the former and decreasing trend in the case of the latter. This near constancy may possibly be due to the fact that the social networks of agents can already be quite complex at $c = 0$ when $\gamma_0$ is large enough, meaning that there is no clear transition from one type of a network to another when $c$ is increased in this case, at least no transition that shows up in the susceptibility and path lengths. 

As noted above, the assortativity and homophily coefficients tend to fluctuate very strongly in time as the simulations progress, often changing even signs. Nevertheless, taking averages over these quantities reveals some interesting details on their behaviour as functions of the model parameters. For example, both the homophily and assortativity coefficients have increasing trends as functions of $c$ when $\gamma_0 = 0$ and generally decreasing trends otherwise, with only slight exceptions. The values of assortativity coefficient are limited approximately to the interval between $-0.2$ and $0.2$ when $\gamma_0 \neq 0$, and rise to little over $0.5$ when $\gamma_0 = 0$. In contrast, the values of the homophily coefficient stay mostly positive for all values of $c$ and $\gamma_0$, except for a point at $c = 0.5$, when $\gamma_0 = 0$. Thus one can draw the conclusion that, while the agents do not have a clear preference on whether to connect to similarly connected agents or not in the most cases studied, they do slightly favour forming connections to agents with similar accumulated monetary reserves.
 
Other than the $\gamma_0 = 0$ case, Fig. \ref{ultcfig} contains relatively little information on the effects of the parameter $\gamma_0$. While quantitative differences between results obtained using different nonzero values for $\gamma_0$ exist, they tend to be small. Some systematic trends can be discerned, however. For example, the results for maximum cluster size and average numbers of clusters start at similar levels for all nonzero values of $\gamma_0$, but they tend to drift apart as $c$ increases. The most striking effects, however, are seen in the average numbers of neighbours and clustering coefficient, both of which clearly increase as functions of both $\gamma_0$ and $c$, a trend that is almost linear for the former quantity. It is remarkable that for $c = 0$, average number of neighbours seems to follow the value of $\gamma_0$ almost precisely. 
 
\begin{figure}[h]
\begin{center}
\epsfxsize = 0.73\columnwidth \epsffile{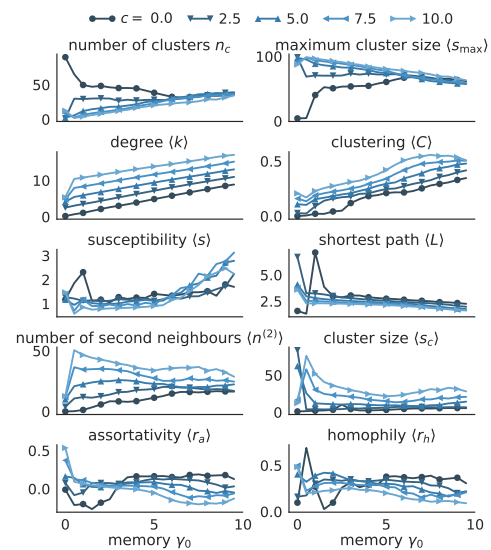}
\caption{The network properties as functions of $\gamma_0$. The linestyles indicate, which of the five constant values of $c$ is being shown.}
\label{ultgfig}
\end{center}
\end{figure}

In order to study the effects of the $\gamma_0$ parameter more closely, we repeated the same exercise as we did for parameter $c$, and calculated the averaged network properties as functions of the parameter $\gamma_0$, for $c = 0, 2.5, 5, 7.5, 10$. The results are shown in Fig. \ref{ultgfig}. Just by looking at the figure we can see similar general issues as noted in the case of Fig. \ref{ultcfig}, i.e. the $c = 0$ case often behaves differently to the others, while the results obtained with other values of $c$ often resemble each other qualitatively, with relatively small quantitative differences. In the cases of the average numbers of clusters, maximum cluster size and path length, for instance, the results tend to converge to very similar values and trends for all $c$ as $\gamma_0$ is sufficiently high, about $6$ in the case of the average numbers of clusters and maximum cluster size, and about $2$ in the case of the average path length. Even below these thresholds the results tend to match for $c \geq 5$, while the $c = 0$ and $c = 2.5$ cases tend to deviate from the others. 

The general trends are slowly decreasing for the path length and the maximum cluster size, while the average number of clusters has a generally slowly increasing trend. The results for the $c = 0$ case show trends opposite to those of the other cases for low values of $\gamma_0$, and the changes tend to be more drastic. It is interesting to note that, while the numbers of clusters creeps up and the maximum cluster size creeps down, the average cluster sizes become almost constant for all $c$ values tested, when $\gamma_0$ is sufficiently large. Overall, it seems that the parameter $\gamma_0$ encourages the formation of small splinter groups outside the main component of the social network, thus the results on the cluster numbers and their maximum sizes.

The results for the susceptibility and path length shown in Fig. \ref{ultgfig} exhibit a similar feature identified in Fig. \ref{ultcfig}, that is, a spiking behaviour at $\gamma_0 = 1$ when $c = 0$, which is related to the formation of the long chains of agents and their linking together, as discussed earlier. A key difference is that the spikes in the susceptibility and path length are not nearly as prominent as functions of $\gamma_0$ as they are as functions of $c$. Other than the spike, the results for susceptibility tend to acquire rather similar values for all $c$ and $\gamma_0$, although the they drift apart somewhat as $c$ becomes large. The general trend is increasing, reflecting the increasing numbers of separate clusters as $\gamma_0$ rises.

The average numbers of neighbours and the clustering coefficient behave rather similarly as functions of $\gamma_0$ as they do as functions of $c$. The general trends are increasing, in the case of the former almost linearly so, with the exception of very low values of $\gamma_0$ and high values of $c$. The reason for similar behaviour is again almost surely related to growing trade associations. In the results on the average numbers of second neighbours, however, one can see a clear difference of behaviour between the two cases. While the number of second neighbours tends to increase as a function of $c$, as a function of $\gamma_0$ it is only rising almost monotonically when $c = 0$. For all the other values of $c$ tested it first rises to some maximum point and then starts decreasing ever more slowly, eventually becoming essentially constant. The steepness of the increase and the decrease, along with the point where the maximum are obtained, depend on the value of $c$ tested: For example, for $c = 7.5$ and $c = 10$ the rise is very steep, the maximum occurs at $\gamma_0 = 0.5$ and the decline is also relatively rapid. For $c = 2.5$ and $c = 5.0$, in contrast, the rise is rather slow, the maximums occur at $\gamma_0 = 6$ and at $\gamma_0 = 3$, respectively, and the decline is almost imperceptible. The reason limiting the rise of the number of second neighbours may be related to the proliferation of splinter clusters, especially of cartels, since they tend to be fully connected.

The behaviours of the average assortativity and homophily coefficients as functions of $\gamma_0$ is characterised by relatively slow changes, with the exception of the $c = 0$ case. While the total change over the full range of $\gamma_0$ can be significant in some cases, minimal changes from a neighbouring value of $\gamma_0$ are the rule. Overall, the assortativity and homophily coefficients acquire similar values as functions of $\gamma_0$ as they do as functions of $c$, and therefore similar conclusions apply. 

\begin{figure}[h]
\begin{center}
\epsfxsize = 0.73\columnwidth \epsffile{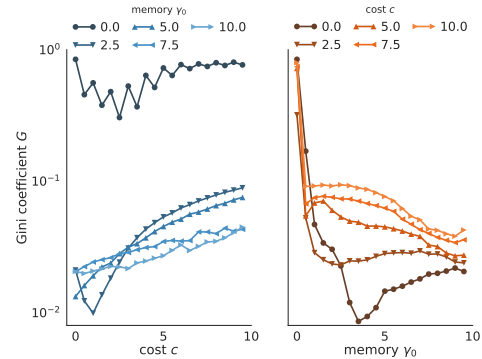}
\caption{Gini coefficient $G$ as function of $c$ (left panel) and $\gamma_0$ (right panel). The linestyles indicate, which of the five constant values of $\gamma_0$ or $c$ is being shown.}
\label{ultgini}
\end{center}
\end{figure}

The wealth distribution, as measured by the gini coefficient, is shown in Fig. \ref{ultgini}. Generally, the gini coefficient decreases as function of $\gamma_0$ and increases as a function of $c$, except in the $\gamma_0 = 0$ case, which is characterised by sharp fluctuations as a function of $c$. Also, in when  $c = 0$ or $c = 2.5$, the gini coefficient starts slowly rising after steep decrease as a function of $\gamma_0$, contrary to the general trend. When $\gamma_0 \neq 0$, one sees that smaller the $\gamma_0$, the greater the speed of the rise of the gini coefficient is as a function of $c$. Conversely, the greater the value of $c$, the slower the decline of the gini coefficient will be as a function of $\gamma_0$. It can be thus concluded that the parameter $c$ generally increases wealth disparities between the agents, while $\gamma_0$ tends to decrease them for a most part, at least to a point.

\subsection{The effects of the population number}

\begin{figure}[h]
\begin{center}
\epsfxsize = 0.73\columnwidth \epsffile{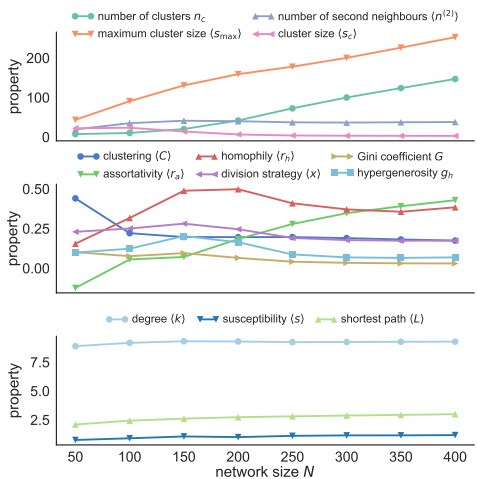}
\caption{Various network properties as functions of the total population $N$, when $c = 7.5$ and $\gamma_0 = 2$. The upper panel shows the average numbers of clusters, second neighbours and cluster size along with maximum cluster size on average. The middle panel shows the average division strategies and hypergenerosity prevalence along with average gini, assortativity, homophily and clustering coefficients. The lowest panel shows the average path length, susceptibility and average number of neighbours. }
\label{ultpop}
\end{center}
\end{figure}

Most of the simulations in this paper have only had $100$ agents due to time and computational constraints. To test the behaviour of our model with different numbers of agents, we performed simulations with $N = 50, 100, 150, 200, 250, 300, 350$ and $N = 400$ agents, with parameter values $c = 7.5$ and $\gamma_0 = 2$, which were chosen for their ability to bring the average division strategies and hypergenerosity prevalence close to those observed at $N = 100$, as shown above. The results are shown in Fig. \ref{ultpop}. One of the more striking revelations from this Figure is that the population number seems to have relatively little effect on many of the results of the model, especially the average number of neighbours, path length and susceptibility, which only show very slight increasing trends. The gini coefficient also changes little as a function of the population number, although its general trend is a very slowly decreasing.

Most important of the results not affected much by the population numbers are the average division strategies and hypergenerosity prevalence. The average division strategy show a very slight growing trend when $N \lesssim 150$, after which there is a slow decreasing trend until $N = 400$, at which point the division strategies have fallen slightly below $0.2$ on average. Hypergenerosity prevalence follows generally the very same pattern, peaking at a value of about $0.2$ when $N = 150$ and slowly decreasing afterwards to a value of less than $0.1$ at $N = 400$. While the average division strategy and hypergenerosity prevalence obtained from the simulations clearly correspond to the observations best at $N = 100$, which is the point at which the calibration of the model parameters was made, the fact that they stay relatively close to the observations raises hopes for the general applicability of the model. In this study we do not, however, venture beyond $N = 400$ in our investigations, so we cannot say exactly how the system behaves at very large population numbers. 

Of the network properties shown Fig. \ref{ultpop} the maximum cluster size, average number of clusters, and assortativity coefficient exhibit the greatest changes and most systematic trends, all of which happen to be increasing, while the average clustering coefficient has a general decreasing trend, which is rapid at first but slows down considerably when $N \geq 100$. In the upper panel of Fig. \ref{ultpop} we see an interesting linkage between the average numbers of clusters and maximum and average cluster sizes. While the increase in maximum cluster size as a function of the population number is a matter of course, the simultaneous strong rise in the average number of clusters drags the cluster sizes down. This effect is seen both in the maximum and average cluster sizes: Although the maximum cluster size is very close to $N$ when $N \lesssim 150$, the relative gap between $N$ and the maximum cluster size gradually widen as $N$ increases, and so at $N = 400$ the maximum cluster size is only about $250$. The average cluster size, however, shows a near consistent downward trend, which is necessarily due to the large number of clusters generated by the model at ever larger population numbers.

The rising trend in the assortativity coefficient reveals the changing preferences of the modeled agents in regarding relation formation. While the social networks are dissociative at $N = 50$,  they become increasingly more associative at higher population numbers. The homophily coefficient does not share such a straightforward trend, as it is at times increasing and at other times decreasing, but always positive, meaning that the agents will always favour forging or keeping ties to other agents with similar accumulated wealths. It should be emphasized, however, that both these coefficients, and especially the assortativity coefficient, are subject to very strong temporal fluctuations during the simulations, so these effects are only present in the average sense.

In summary it could be said that the results on the cluster sizes and numbers shown in Fig. \ref{ultpop} indicate that the model produces ever greater amounts of ever smaller clusters that splinter off the main component as population numbers increase, while the decreasing clustering coefficient and the increasing assortativity coefficient indicate that the size of the patron-client networks grows relative to the size of the trade associations. It should be noted, however, that these results have been obtained using only one set of model parameters calibrated at $N = 100$ to replicate the observed results. We do not delve deeper into the interaction of the model parameters and the population number in this study.

\section{Conclusions and Discussion}
\label{con}

In this study  we have investigated the behaviour of a network of agents seeking to maximize their relative standings, according to better than hypothesis (BTH). The agents are embedded in a co-evolving network, in which the linked agents repeatedly play the dictator game with each other for evolving their social relations, while their status is measured by the amount of wealth they thus acquire. The main motivation of this research is to test, whether the agents driven by BTH would form any connections or endow anything to their network partners. In evolving their social relations the agents in this network game keep track of how the other agents treat them forming social relations but also cutting them as punishment of selfish behaviour, and "forget" their treatment as well as spend a fixed portion of their earnings for paying their living costs once per game cycle. The cost of living and the rate at which the agents ``forget'' their treatment by the other agents are parameters of our model, and we have studied their influence to the behaviour of the system of agents. 

Our simulation results show that agents acting along the BTH do, indeed, form social connections for the purpose of playing the dictator game and that the dictators in these games often give non-zero amounts of money to the other players. That the agents would give each other anything at all in a game such as the dictator game is not a self-evident conclusion, a priori. The agents can use either very generous or very stingy strategies depending on the model parameters and their position in the co-evolving network. 

Generally speaking, the structures of the social networks produced by the model vary strongly according to the model parameters, and can be described as follows. For small values of the model parameters the system networks reduce to collections of short chains of agents, which become longer as either model parameter is increased. Ultimately these chains start fusing together at high values of the parameters. The cost parameter alone does not seem to have as dramatic effect on the structure of the networks as the memory parameter, but it makes the agents form more connections, especially in conjunction of non-zero memory parameter, thereby making the network denser.

When the parameters are sufficiently large new substructures, which we name "trade associations" and "patron-client networks" start emerging. The former are fully connected subgroups of agents that use relatively similar division strategies, while the latter are composed of agents that have diverse division strategies and relatively sparse social connections, with generous agents generally connecting to many stingier agents that in turn do not form many connections, and virtually none with each other. The emergence of these substructures demonstrates on one hand that the agents driven by BTH are capable of forming complex social structures using diverse strategies, and on the other hand that they can at some level form social norms. Especially the fact that the members of trade associations coordinate their division strategies, indicates some appreciation of fairness in part of the agents. 

The results of our dictator game of networked agents are in agreement with the empirically findings of altruistic behaviour by humans in the role of the dictator, which provides credence to the BTH (Better Than Hypothesis). We find it very interesting that, based on rather simple assumptions about the competition for superior social positions, the dynamics generates complex network structures indicating that this component of human behaviour may have an important role in producing the empirically observed structures in real societies.
It is also notable that with suitable parameter values our model produces average hyper-generosity prevalence and division strategies that are reasonably in line with the ones observed in earlier research. That BTH could mimic these observational facts, on top of being capable of facilitating formation of complex social structures even at such a simplistic level, is encouraging when considering possible future uses of the BTH framework. However, it should be emphasised that as such our model parameters do not  correspond to anything directly observable. In the future work, parameters akin to the memory parameter could, for example, be replaced with more detailed  social mechanisms such as giving social relationships a value of their own, and letting simulated agents compete over them. 

\section*{Funding}

This work was supported by Niilo Helander's foundation grant No. 160095 (J.E.S.), the Academy of Finland Research project (COSDYN) No. 276439 and EU HORIZON 2020 FET Open RIA project (IBSEN) No. 662725 (K.K.), and Conacyt projects 799616 and 28327 (R.A.B.). R.A.B is also grateful for a sabbatical grant from PASPA, DGAPA, UNAM, Mexico.

\section*{Acknowledgments}

Computational resources provided by the Aalto Science-IT project have been utilised in this work.  G.I. and J.K. thank Aalto University for hospitality.

\bibliographystyle{unsrt}
\bibliography{BTHgeneral}

\end{document}